\documentclass[a4paper,usenatbib]{mnras}
\usepackage{CJK}
\usepackage{hyperref}
\usepackage{mathtools} 
\usepackage{amssymb}
\usepackage{booktabs}
\hypersetup{
  pdftitle		={Heat Pumps},
  pdfauthor		={Yi-Hao Chen},
  colorlinks	= true, 
  urlcolor		= blue, 
  linkcolor		= blue, 
  citecolor		= blue  
}

\newcommand{\code}[1]{\texttt{#1}}
\newcommand*\dd{\mathop{}\!\mathrm{d}}

\title[Heat Pumps in Galaxy Clusters]{Jets, Bubbles, and Heat Pumps in Galaxy Clusters}

\author[Y.-H. Chen et al.]{
	Yi-Hao Chen,$^{1}$\thanks{E-mail: ychen@astro.wisc.edu}
	Sebastian Heinz,$^{1}$
	and Torsten A. En{\ss}lin$^{2}$
\\
$^{1}$Department of Astronomy, University of Wisconsin-Madison,
475 N. Charter Street, Madison, WI 53706, USA\\
$^{2}$Max Planck Institut f\"{u}r Astrophysik, Karl-Schwarzschild-Str 1, D-85740 Garching, Germany
}

\date{Accepted XXX. Received YYY; in original form ZZZ}

\pubyear{2019}

\begin{document}
\label{firstpage}
\pagerange{\pageref{firstpage}--\pageref{lastpage}}
\maketitle

\begin{abstract}

  Feedback from AGN jets has been proposed to counteract the catastrophic
  cooling in many galaxy clusters. However, it is still unclear which physical
  processes are acting to couple the energy from the bi-directional jets to the
  ICM. We study the long-term evolution of rising bubbles that were inflated by
  AGN jets using MHD simulations. In the wake of the rising bubbles, a
  significant amount of low-entropy gas is brought into contact with the hot
  cluster gas. We assess the energy budget of the uplifted gas and find it
  comparable to the total energy injected by the jets. Although our simulation
  does not include explicit thermal conduction, we find that, for reasonable
  assumptions about the conduction coefficient, the rate is fast enough that
  much of the uplifted gas may be thermalized before it sinks back to the core.
  Thus, we propose that the AGN can act like a geothermal \emph{heat pump} to
  move low-entropy gas from the cluster core to the heat reservoir and will be
  able to \emph{heat} the inner cluster more efficiently than would be possible
  by direct energy transfer from jets alone. We show that the maximum
  efficiency of this mechanism, i.e. the ratio between the conductive thermal
  energy and the work needed to lift the gas, $\xi_{\mathrm{max}}$ can exceed
  100 per cent. While $\xi < \xi_{\mathrm{max}}$ in realistic scenarios,
  AGN-induced thermal conduction has the potential to significantly increase
  the efficiency with which AGN can heat cool-core clusters and transform the
  bursty AGN activities into a smoother and enduring heating process.

\end{abstract}


\section{Introduction}

The cores of galaxy clusters often exhibit strong x-ray emission with cooling
times short compared to the cluster age, which should lead to substantial
radiative cooling and the formation of a \emph{cooling flow}
\citep{Fabian1994}. The lack of cold gas in observations \citep{Peterson2003,
Peterson2006} motivates the study of feedback processes that could counteract
the cooling \citep{McNamara2007, Fabian2012}. Mechanical energy deposited by
AGN jets could serve as the mechanism to regulate the thermal evolution of
galaxy clusters \citep[see][for review]{McNamara2012}. Strong evidence
supporting this mechanism includes X-ray observations of galaxy clusters, which
often show radio-filled cavities that are likely inflated by jets from the
supermassive black holes in the central galaxy \citep[see e.g.][]{Birzan2004}.
In many cases, multiple cavities, likely caused by episodic jet activity, are
observed. The most detailed observations of this kind include the Perseus
Cluster \citep{Fabian2011} and M87 \citep{Forman2007}. \cite{Churazov2000}
show that AGN jets have comparable power to offset the cooling in the Perseus
Cluster. Observations of numerous other cool core clusters support this claim,
but with large scatter in jet power \citep[see e.g.][]{Rafferty2006,
McNamara2007, Hardcastle2019}.

However, the processes through which the energy of AGN jets couples to the ICM
are still uncertain. The highly directional nature of jets poses substantial
difficulty to distribute their energy isotropically in the core of galaxy
clusters \citep{ONeill2010}. \cite{Fabian2017} argue that the dissipation of
sound waves, which are visible as ripples in the x-ray observations, could heat
the ICM. \cite{Heinz2006} and \cite{Morsony2010} show that large-scale motion
of the cluster helps to disperse the energy from the AGN. Other possible
mechanisms include turbulence \citep{Zhuravleva2014}, mixing \citep{Hillel2016,
Yang2016, Hillel2018}, shocks \citep{Fabian2006, Li2017}, internal waves
\citep{Zhang2018}, and cosmic ray streaming \citep{Ensslin2011, Ehlert2018}
among others.

Although most of the discussion of heating focuses on the direct energy
coupling between the AGN output and the ICM, a few authors have alluded
that the removal of the cool gas from the core of the cluster could possibly
prevent catastrophic cooling. \cite{Yang2016a} point out that a ``gentle
circulation`` can help isotropize the heating from the weak shock and mixing
that happens primarily in the jet cone. Uplifted gas trailing the bubbles was
also noticed in simulations \citep{Weinberger2017, Duan2018}. \cite{Pope2010}
calculate the amount of gas transported by a rising bubble and conclude that
the mass transport by the bubble wake could prevent the core from overcooling.

Buoyantly rising bubbles are identified and studied by many works.
\cite{Churazov2001} conduct hydrodynamic simulations of bubbles to model the
radio and x-ray arms in M87. In a deep \textit{Chandra} observation,
\cite{Forman2007} detected many filamentary structures that could be associated
with independent bubbles in M87. \cite{Wise2007} identify a series of cavities
in the Hydra A cluster. \cite{Gendron-Marsolais2017} report observations of
cool x-ray gas rims in NGC 4472 that could be gas lifted by the bubbles.
\cite{Su2017} find the gas uplift rate is comparable to that expected to cool
in low-mass Fornax cluster. These works all suggest that rising bubbles are
prevalent and could bring gas from the core to the outskirt of the galaxy
cluster.

Although thermal conduction is considered insufficient to offset the cooling in
cool-core clusters by itself \citep{Voigt2004, Dolag2004}, it still plays a
vital role in the thermal state of the cluster, especially when perturbed by
the rising bubble. In this work, we propose and investigate the effect of
rising bubbles bringing low-entropy gas from the cool core to the hot outskirt
of the galaxy cluster where thermal conduction is much more efficient. With
this mechanism, the AGN could ``heat'' the ICM with an efficiency that can
exceed 100 per cent, drawing on the excess heat contained in the outer cluster
like a heat pump. This heating process is also more gentle than direct heating.

This work is organized as follows. We describe the methodology and techniques
in Section \ref{sec:numerical_techniques}. In Section \ref{sec:results}, we
present the results of the simulations, including the energy budget and the
conduction rate estimates. We derive and calculate the efficiency of this
mechanism in simplified profiles in Section \ref{sec:efficiency}. In Section
\ref{sec:discussion} and \ref{sec:conclusions}, we discuss and summarize our
findings.

\section{Numerical Setup and Techniques}
\label{sec:numerical_techniques}

Here we describe the important and relevant details of the simulation. We
conduct full 3D ideal MHD simulations using \code{FLASH} \citep{Fryxell2000,
Dubey2009} with the unsplit staggered mesh scheme \citep{Lee2013} and AMR. We
simulate the cluster as a spherically symmetric environment, tuned to match the
properties of the cool-core Perseus Cluster following \cite{Zhuravleva2015}.
The density profile follows a $\beta$-model

\begin{equation}
	\label{eq:beta_model}
	\rho(r) = \rho_0/[1+(\frac{r}{r_c})^2]^{\frac{3}{2}\beta},
\end{equation}
with $\rho_0 = 9.6\times10^{-26}$ g cm$^{-3}$, $r_c = 26$ kpc, $\beta = 0.53$.
The temperature profile is set such that the core of the cluster is cooler than
the outskirts,
$T(r) = T_{\mathrm{out}} \left( [1+(\frac{r}{r_{\mathrm{c,T}}})^3] \right)
/ \left( \frac{T_{\mathrm{out}}}{T_{\mathrm{core}}}+
  (\frac{r}{r_{\mathrm{c,T}}})^3 \right)$, with $T_{\mathrm{core}} = 3.0$
keV, $T_{\mathrm{out}} = 6.4$ keV and $r_{\mathrm{c,T}} = 60$ kpc.
We assume monatomic gas in the cluster so that the adiabatic index $\gamma$ =
5/3 and choose the mean molecular weight $\mu$ = 0.61. The cluster
potential is assumed to be spherically symmetric and static throughout
the simulations, neglecting changes in the gravitational potential due to the
changes in gas density. The fixed gravitational potential is chosen so that
the initial conditions are in hydrostatic equilibrium. Our simulations were
set up to study a wide context of questions relating to jet and cluster physics
that do not require magnetization of the ICM, so it was not included in the
initial conditions of our simulations. The simulation box is $1 \times 0.5
\times 0.5$ Mpc with hydrostatic diode boundary condition
(\code{hydrostatic-f2+nvdiode} in \code{FLASH}). The adaptive refinement
criteria include the native second derivative based estimator and also a
self-defined momentum-based condition that ensures that we resolve the jets at
the highest resolution (30 pc). The jets are active at a constant jet power
until 10 Myr. We restrict the maximum refinement level for the ICM further
away from the center. To study the motion of the gas, we use Lagrangian tracer
particles in \code{FLASH} \citep{Dubey2012}. Using density-weighting, we
randomly distribute $5\times 10^{5}$ particles within a 150 kpc radius in the
ICM. Thus, each particle represents approximately the same mass.

We set up a \emph{nozzle} through which energy, momentum, and magnetic
flux are injected into the simulation grid to model the jets from the
central AGN. The magnetic pressure is in equilibrium with the thermal pressure
in the nozzle, i.e. plasma $\beta$=1. We further set the jet power ($10^{45}$
erg s$^{-1}$), jet speed (0.1c), nozzle radius (240 pc), and the internal Mach
number (10), which sets the jet density. The magnetic field of the jet nozzle
is set to be poloidal, i.e., parallel to the jet propagation
direction\footnote{We also performed simulations with other topologies which
are fully consistent with the results shown here, see Chen et al. in prep.}.
The jet is active for 10 Myr, during which it operates at constant power. The
direction of the jet is set up to have a small angle ($\sim$ 8 degrees)
jittering to mimic the dentist's drill effect seen in observations and inferred
theoretically \citep{Scheuer1982}. After 10 Myr, the jet shuts off and the
bubbles inflated by the jets continue to rise. For computational expediency,
the highest resolution of the simulation is reduced to 120 pc shortly after
turning off the jet (from 30 pc when the jet is active). Radiative cooling and
explicit thermal conduction are not included in the simulation. We investigate
the long-term impact of this one-time activity of the jets on the dynamical
state of the galaxy cluster. Visualizations and profile analysis are performed
with the \code{python}-based software package
\code{yt}\footnote{\url{http://yt-project.org/}} \citep{Turk2010}.

\section{Results}
\label{sec:results}
\subsection{Motion of the low-entropy gas}

\begin{figure*}
\begin{minipage}{1\textwidth}
	\includegraphics[height=9.5cm]{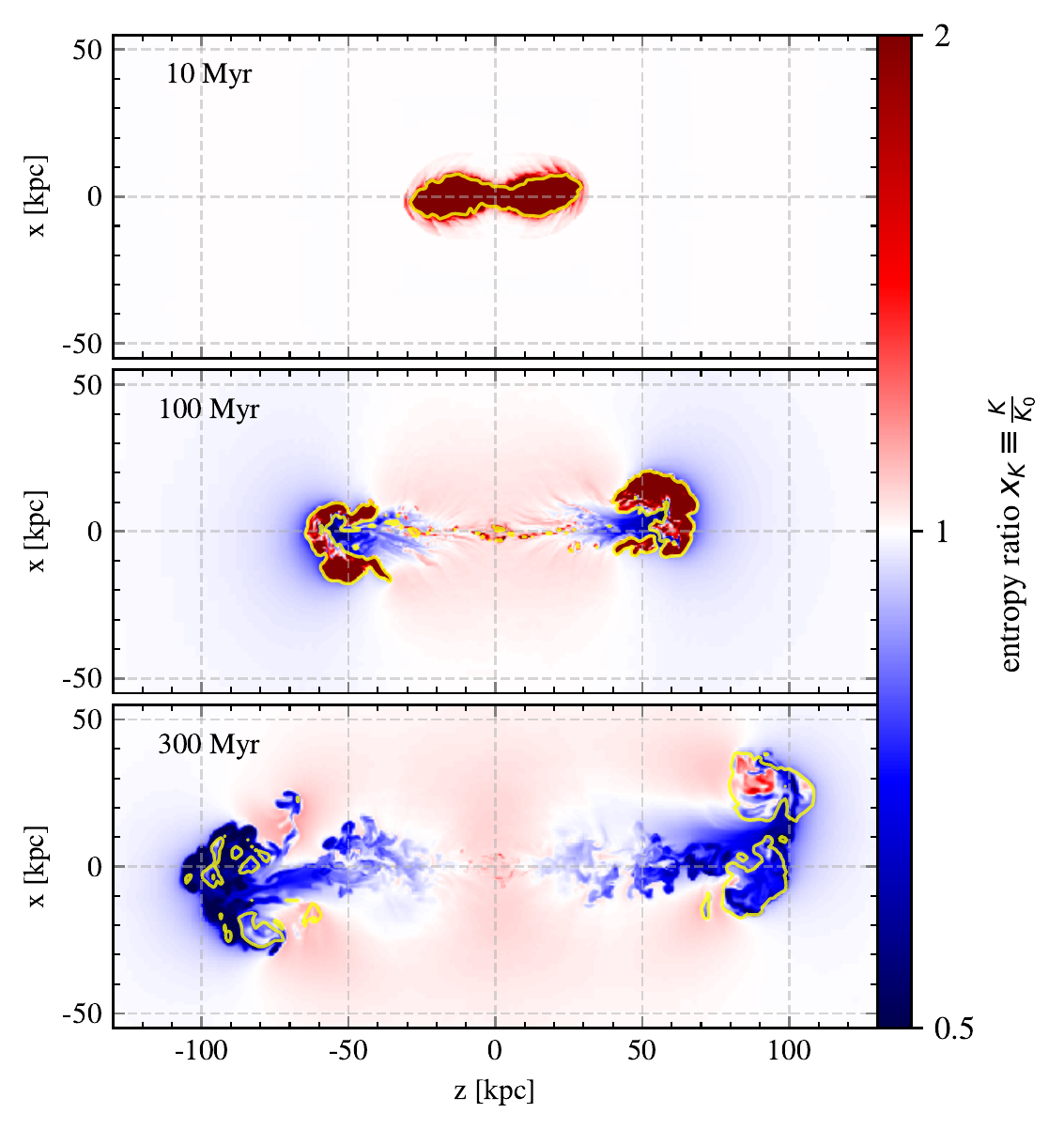}
	\includegraphics[height=9.5cm]{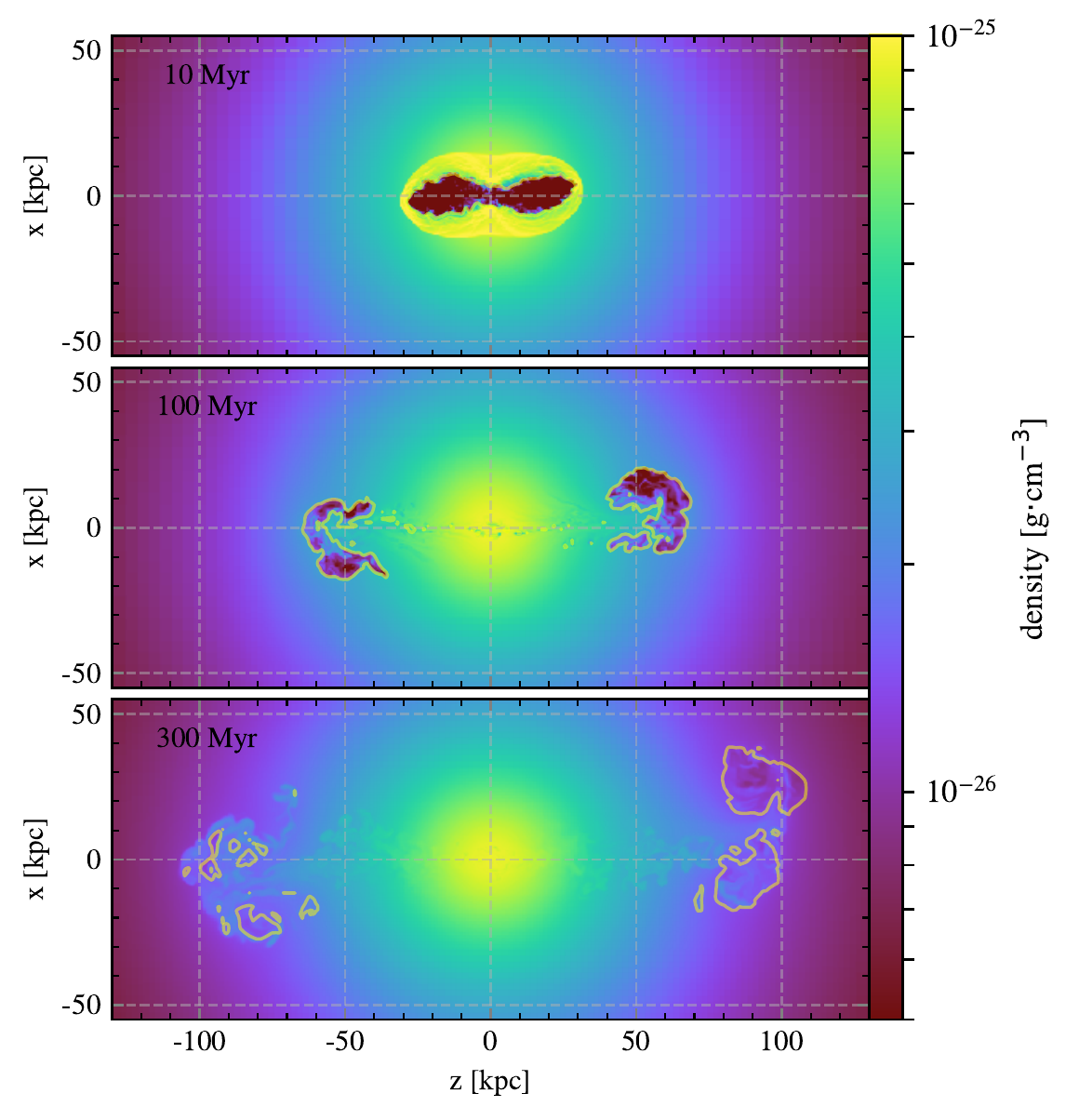}
	\caption{Central slices of the simulation showing (\emph{left}) the entropy
		of the gas relative to the initial entropy value at the same location
		and (\emph{right}) the density of the gas at different times in the
		simulation. We can see the bubbles lift the low-entropy gas, while the
		cluster core is refilled by the higher-entropy gas. Yellow contours
		indicate the jet mass fraction of $10^{-3}$, above which the conduction
		rate is excluded in the calculation in Section
		\ref{subsec:spitzer_heating_rates}. }
	\label{fig:entropy_ratio}
\end{minipage}
\end{figure*}

First, we investigate the lifting of the low-entropy gas by the rising bubbles.
During the active phase of the AGN, only a small amount of gas is displaced by
the jet. However, as the jet ceases and bubbles rise, they drag a significant
amount of gas from the core in the wake. This phenomenon can be seen clearly in
the entropy ratio maps. In Fig.~\ref{fig:entropy_ratio}, the specific entropy,
defined as $K \equiv kT/n^{\gamma - 1}$, of the gas at different times of the
simulation is plotted relative to the entropy profile of the initial ICM
conditions alongside the density slices. Since the entropy is conserved during
adiabatic processes, the entropy ratio $x_K \equiv K/K_0$ is a good indicator
of the origin of the gas\footnote{We do have shocks in the simulation during
the initial inflation of the cocoon, so some fraction of the gas undergoes
non-adiabatic heating. Its entropy will be correspondingly raised.}. An
entropy ratio smaller than one shows the gas has been lifted radially outward,
while an entropy ratio larger than one could mean that the gas has been moved
inward or reveal the presence of the hot gas from the jet or the shocked
cluster gas. When the jet is still active, at 10 Myr, only the very hot gas
injected by the jets is visible in the entropy ratio map and the initial
entropy profile of the ICM is almost unchanged. This anisotropic heating has
been a challenge for invoking AGN feedback to counteract the strong x-ray
cooling \citep{ONeill2010}.

However, the bubbles keep rising due to both buoyancy and momentum injected by
the jets. At 100 Myr (middle panel of Fig.~\ref{fig:entropy_ratio}), we can see
there is lower-entropy gas pushed ahead of the bubbles as well as rising in
their wakes. A slight increase of density can also be seen in the right panels
of Fig.~\ref{fig:entropy_ratio}. However, due to the underlying density
gradient of the cluster profile and the strong contrast between the bubble and
ICM gas, it is not immediately obvious except in the bottom panel. The sphere
inside the location of the bubble ($\lesssim$ 50 kpc) shows a general increase
of entropy because of the replenishment of the gas from larger radii. At 300
Myr (lower panel of Fig.~\ref{fig:entropy_ratio}), most of the hot gas injected
by the jet is mixed--mostly numerically--with the ICM\footnote{It is important
to note that we do \emph{not} count the numerical mixing in our heating rate
later on, as it is a purely numerical artifact.}. The vortex generated by the
bubble brings a notable amount of gas moving outward to large radii even
further than the bubble. Only a small fraction of the hot gas is still visible
as a ring surrounded by the lower-entropy gas. If there is no heat exchange
between the high- and low-entropy gas, the low-entropy gas will eventually fall
back to the core again. In Section \ref{subsec:spitzer_heating_rates}, we argue
that the conduction rate will increase when the low-entropy gas is brought into
contact with the hot atmosphere, thus heating the gas efficiently before it
sinks.

\subsection{Percentage of gas lifted by the rising bubbles}

\begin{figure}
\begin{center}
	\includegraphics[width=0.49\textwidth]{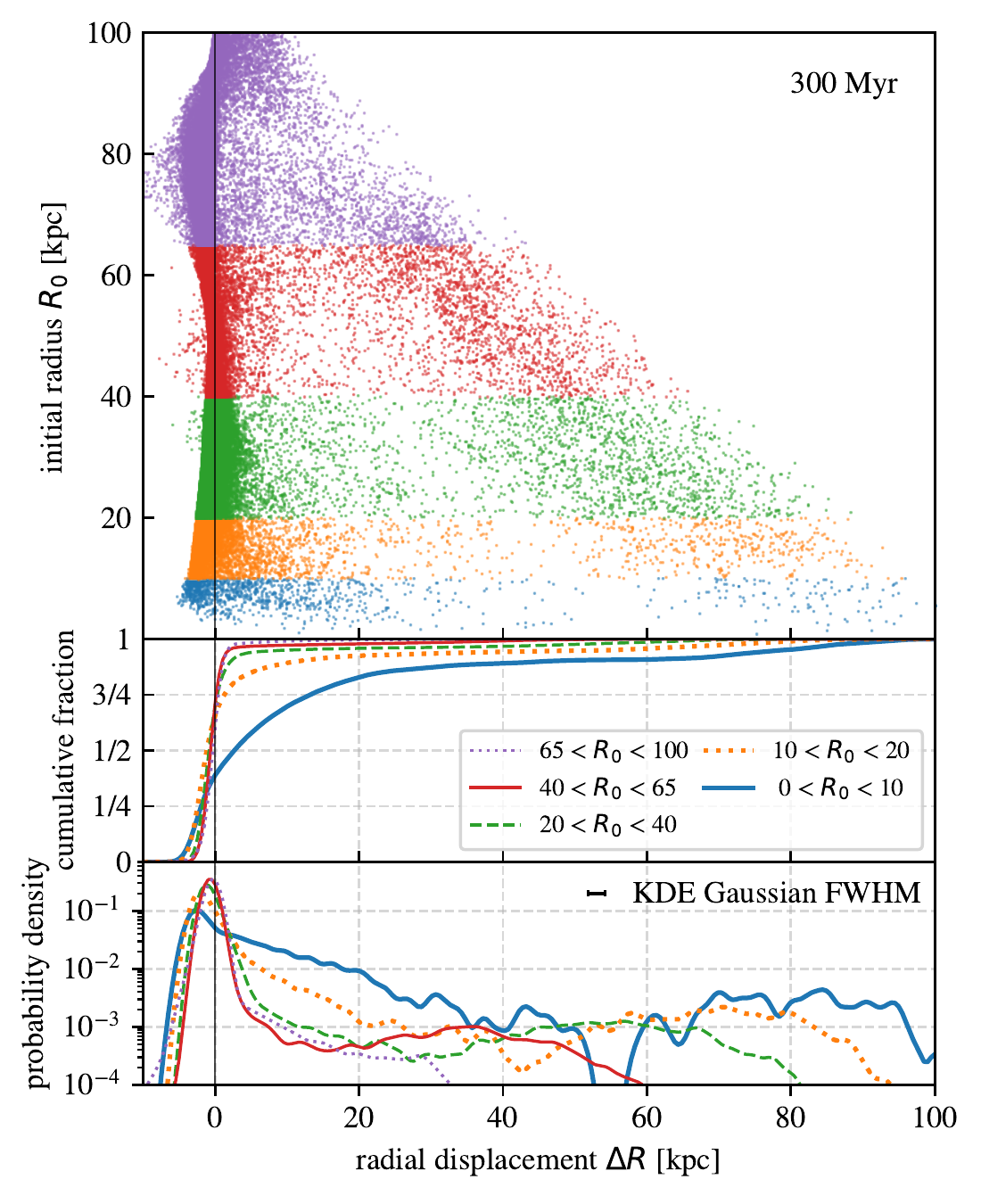}
	\caption{Displacement of tracer particles at 300 Myr. The change in radius
		(horizontal axis) is plotted against the initial radius (vertical axis,
		upper panel) of the tracer particles. The particles are initialized
		such that the number of particles is proportional to the density of the
		gas. The middle and lower panel shows the cumulative distribution and
		probability density function using kernel density estimation for the
		particles grouped by their initial radii. We can see that around 15 per
		cent of the gas initially within the 10-kpc core is lifted to more than
		20 kpc away. While the low-entropy gas is lifted by the bubbes, rest of
		the higher-entropy cluster gas moves slightly inward to replenish the
		core.}
	\label{fig:particles_dr}
\end{center}
\end{figure}

To quantify the amount of gas being lifted by the rising bubbles, we use tracer
particles to track the movement of gas. The particles are distributed according
to the density of the gas at the beginning of the simulation. Thus, each
particle represents roughly the same mass of gas. The radial displacement of
these particles at 300 Myr is shown in the upper panel of
Fig.~\ref{fig:particles_dr}, in which we group them by their initial radii.

The distribution of the gas displaced by the bubbles can be identified in the
middle and lower panel of Fig.~\ref{fig:particles_dr}. We group the particles
by their initial radii and use the kernel density estimation with the Gaussian
kernel width matching the initial average separation between particles. The
same kernel is used for both the cumulative distribution and probability
density function. The distribution can be viewed as the distribution of mass
for the gas in different initial radii. We can see that a larger fraction of
mass is lifted for the gas closer to the core. For the gas initially inside 10
kpc radius, about 15 per cent of the gas is lifted more than 20 kpc outward;
for the gas in 10 to 20 kpc initial radii, the fraction is slightly less than
10 per cent. This is the effect at 300 Myr of the bubble inflated by the jet
that is active for only 10 Myr. If there were additional episodes of jet
activity, the fraction would be larger.

In Fig.\ref{fig:particles_dr}, we can also see the replenishment of the gas in
the core. The negative $\Delta r$ in both the cumulative fraction and
probability density indicates an inward motion due to the removal of
low-entropy gas from the core. This motion gradually brings the higher-entropy
gas from larger radii to the core. Note that the inflow and the outflow take
place at different regions of the cluster -- the outflow happens in the cone
around the bubbles while the inflow occurs primarily outside the bubble cone.

\subsection{Mass and energy budget of the low-entropy gas lifted by the rising bubble}
\label{subsec:mass_energy_budget}

\begin{figure}
\begin{center}
	\includegraphics[width=0.49\textwidth]{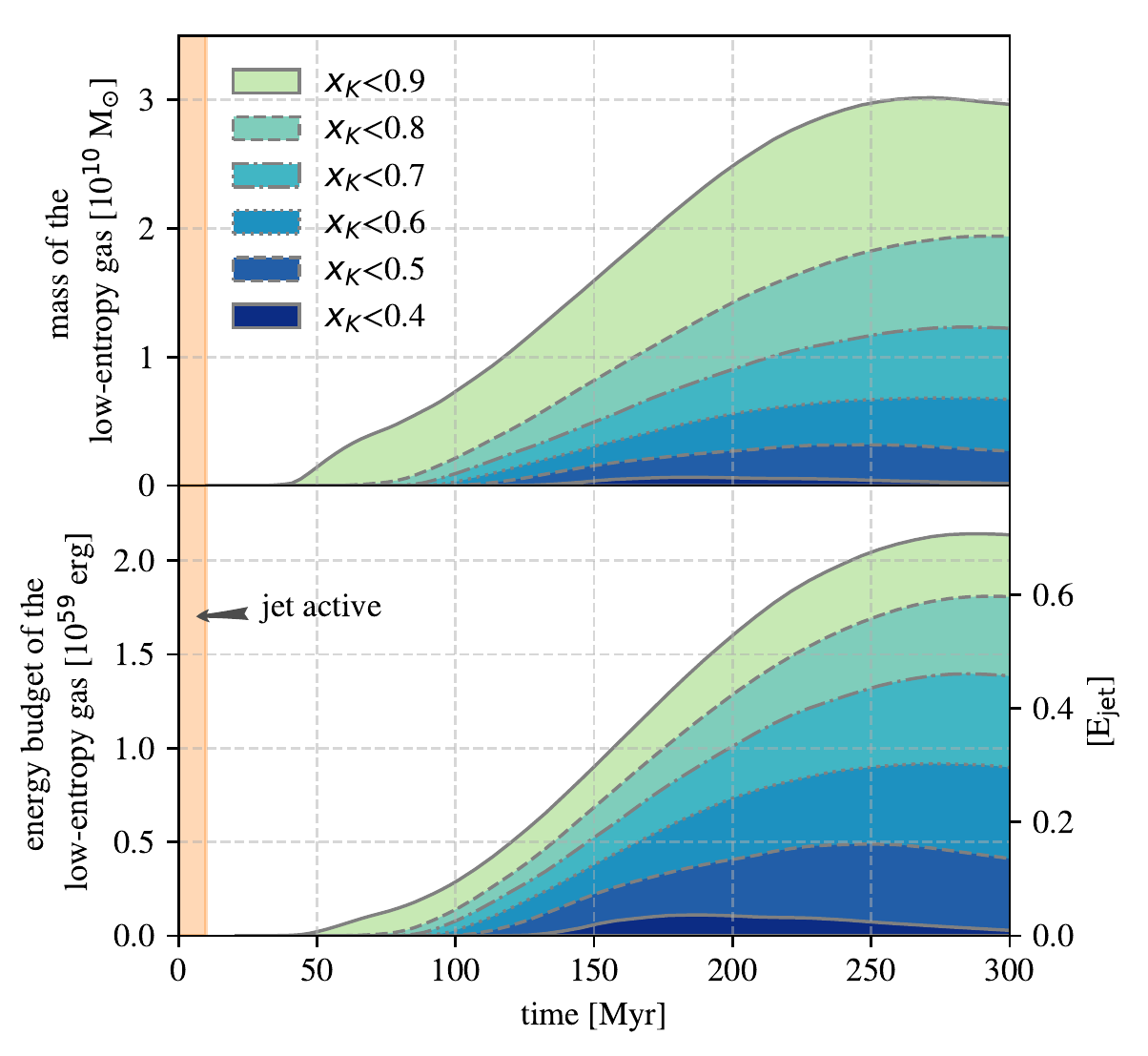}
	\caption{Mass and the energy budget of the low-entropy gas. This figure
		shows (\emph{upper panel}) the evolution of the amount of mass of the
		low-entropy gas, which is identified by entropy ratios $x_K$ lower
		than the threshold, and (\emph{lower panel}) the energy needed to bring
		the gas to the surrounding temperature. The y-axis on the right of the
		lower panel shows the energy budget relative to the total energy
		injected by the active jet in the first 10 Myr.}
	\label{fig:mass_energy_budget}
\end{center}
\end{figure}

We now investigate the mass of the low-entropy gas and the energy that could be
extracted from the hot gas and transferred to the uplifted low-entropy gas by
heat conduction. We identify gas with entropy ratio $x_K$ below various
thresholds in the simulation. The evolution of the mass of the low-entropy is
shown in the upper panel of Fig.~\ref{fig:mass_energy_budget}. We then estimate
the amount of energy needed to bring this gas into thermal equilibrium with its
surroundings, assuming monatomic gas under constant pressure:
\begin{equation}
	E_{\mathrm{th}, x_i} = \int \frac{5}{2} \frac{ k_B \Delta T}{\mu m_p} \dd M_{x_K \le x_i},
\end{equation}
where $\mu$ is the mean molecular weight and $\Delta T$ is the difference
between current and initial temperature of the gas at the same location.

We use this upper limit as a proxy for the amount of thermal energy that could
be absorbed by the uplifted low-entropy gas. The actual energy budget will
obviously depend on the rate of thermal conduction, which will be discussed in
Section \ref{subsec:spitzer_heating_rates}, and the dynamic state of the
cluster. The lower panel of Fig.~\ref{fig:mass_energy_budget} shows the
evolution of this energy as a function of time with different values of $x$.
The vertical axis at the left labels the energy in [ergs], while at the
right it denotes this energy relative to the total energy ($\sim 3.16 \times
10^{59}$ erg) injected during the active phase of the jet (10 Myr).

One can see that the conductive energy budget of the low-entropy gas is
comparable to the total energy injected by the jet and reaches the peak value
at around $t_{\mathrm{peak}}\sim 300$ Myr. The value of $t_{\mathrm{peak}}$
naturally depends on the details of the system, including the density of the
lobe and gravitational potential profile. However, it will likely be on the
order of hundred million years, which is induced by the short active phase (10
Myr) of the jets. Thus, it may transform the bursty activities of the AGN into
a much longer timespan and result in smoother heating of the cluster.

It is imperative to note that the heat exchange occurs between cluster gas, not
with the very high-entropy relativistic plasma inside the bubble, which has a
vanishingly small conduction coefficient. The estimated energy budget is not
directly from the AGN itself, but from the hot atmosphere at large radii of the
clusters, where the cooling time is longer than the Hubble time. When the
cooler gas is brought into thermal contact with the hot reservoir, heat
exchange can take place more efficiently than it could in a stratified cluster
atmosphere. We discuss the rate of heat exchange in the next section.

\subsection{Conductive heating rates}
\label{subsec:spitzer_heating_rates}

Next, we estimate the rate of thermal conduction between the uplifted
low-entropy gas and the hot thermal bath. We consider the classic
Spitzer conduction in ionized gas \citep{Spitzer1962,
  Narayan2001}. The Spitzer coefficient can be calculated as
\begin{equation}
	\label{eq:Spitzer}
	\kappa_{\mathrm{Sp}} \sim 4 \times 10^{32}
	\left( \frac{kT}{10~\mathrm{keV}} \right)^{5/2} \left( \frac{n}{10^{-3}
	\mathrm{cm}^{-3}} \right)^{-1} \mathrm{cm}^2 \mathrm{s}^{-1}.
\end{equation}
The coefficient has a strong dependence on the temperature and, as a
consequence, the conduction rate is slow in the cool core. However, once the
cool gas is lifted to large radii and placed in close proximity to hot gas,
high temperature and increased temperature gradient both accelerate the thermal
conduction rate across the interface. The highly corrugated nature of the
interface (cf. Fig.~\ref{fig:entropy_ratio}) further increases conduction
compared to a stratified isotropic cluster. High-resolution simulations are
critical to resolve this interface.

We then calculate the heat flux by
\begin{equation}
	\vec{q} = n k_B f_{\mathrm{Sp}} \kappa_{\mathrm{Sp}} \nabla T,
\end{equation}
in which $f_{\mathrm{Sp}}$ is the fraction relative to the classic Spitzer
conduction rate and encapsulates various factors including the orientation of
the magnetic fields and the plasma microphysics. Since we do not include
magnetic fields in the initial ICM, it is impossible to include anisotropic
conduction in this analysis. Instead, we conservatively use $f_{\mathrm{Sp}}$
= 0.1 and 0.01, which is much lower than the typical $1/3$ value used in many
studies, and express any findings in terms of $_{\mathrm{Sp}}$, the effective
conductive coefficient in units of the Spitzer value so that results can
readily be understood for a range of assumptions about conductivity.
We further exclude regions where the jet mass fraction exceeds $10^{-3}$ (cf.
contours in Fig.~\ref{fig:entropy_ratio}) to avoid counting the heat exchange
between the relativistic gas in the hot bubbles and the ICM, which is strongly
suppressed. Note that we do not exclude the jet region explicitly in Section
\ref{subsec:mass_energy_budget}, since the selection of the low-entropy gas
naturally does not incorporate the hot jet gas.

\begin{figure}
\begin{center}
	\includegraphics[width=0.49\textwidth]{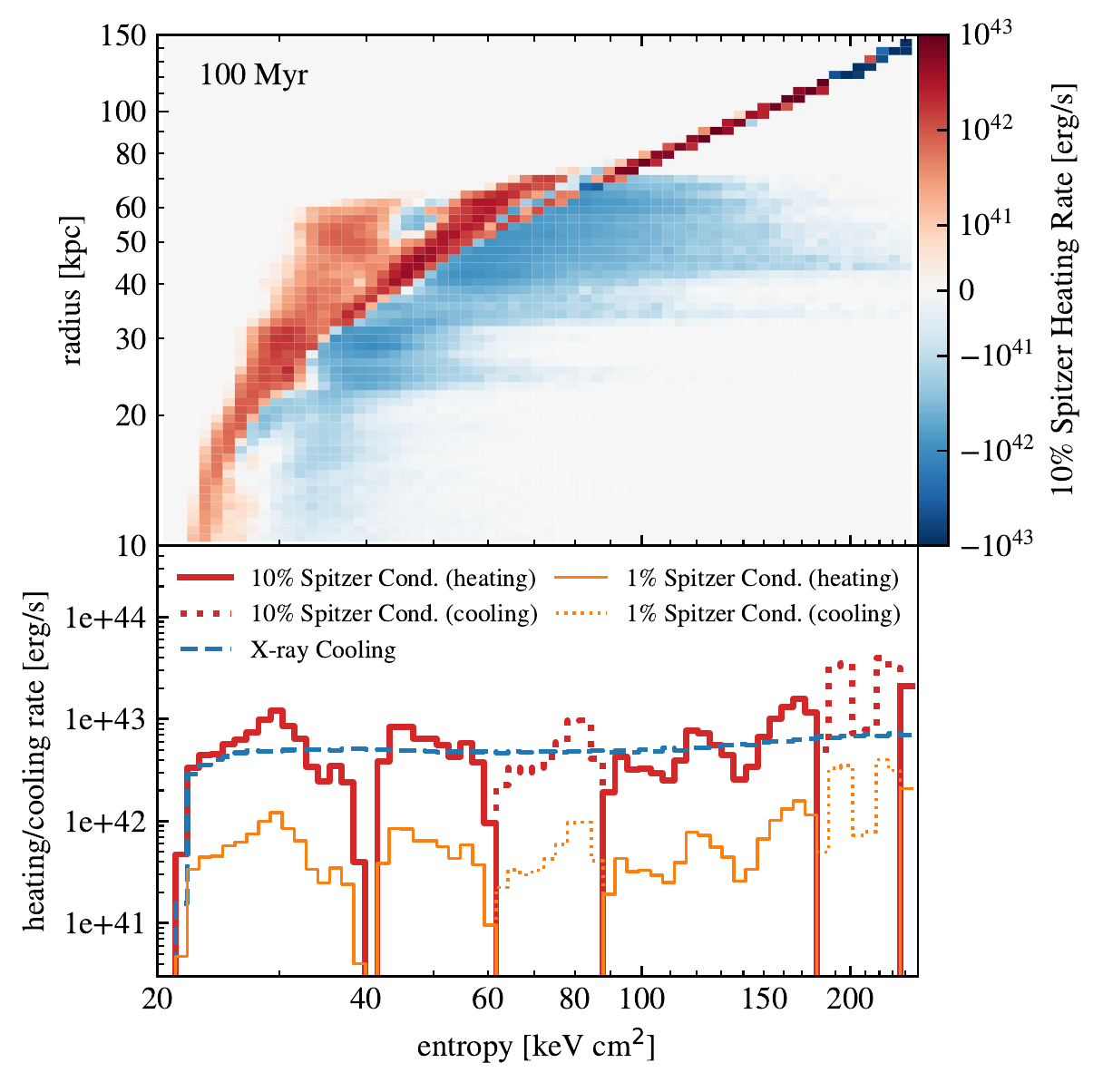}
	\caption{Heating and cooling rate histogram at 100 Myr. The conduction
		heating and cooling rates are summed in each entropy and radial bins
		with $f_{\mathrm{Sp}}=0.1$ (\emph{upper panel}). Marginal histogram
		shows the heating and cooling in entropy bins (\emph{lower panel}).
		For comparison with the x-ray cooling rate, two conduction rates
		$f_{\mathrm{Sp}}$ = 0.1 and 0.01 are plotted. Solid lines indicate
		heating, while dotted and dashed lines denote cooling. Note that the
		conductive heat flux occurs primarily around the surface of the
		uplifted low-entropy gas, while the x-ray cooling happens in all gas.
		The energy is transferred from the high-entropy gas to low-entropy gas
		at the same radius and thus the conductive heating and cooling rates as
		a function of radius have much smaller value. The high-entropy gas that
		cools at small radii ($<$60 kpc) is due to the contamination from the
		hot jet gas.}

	\label{fig:heating_rate_100Myr}
\end{center}
\end{figure}

\begin{figure}
\begin{center}
	\includegraphics[width=0.49\textwidth]{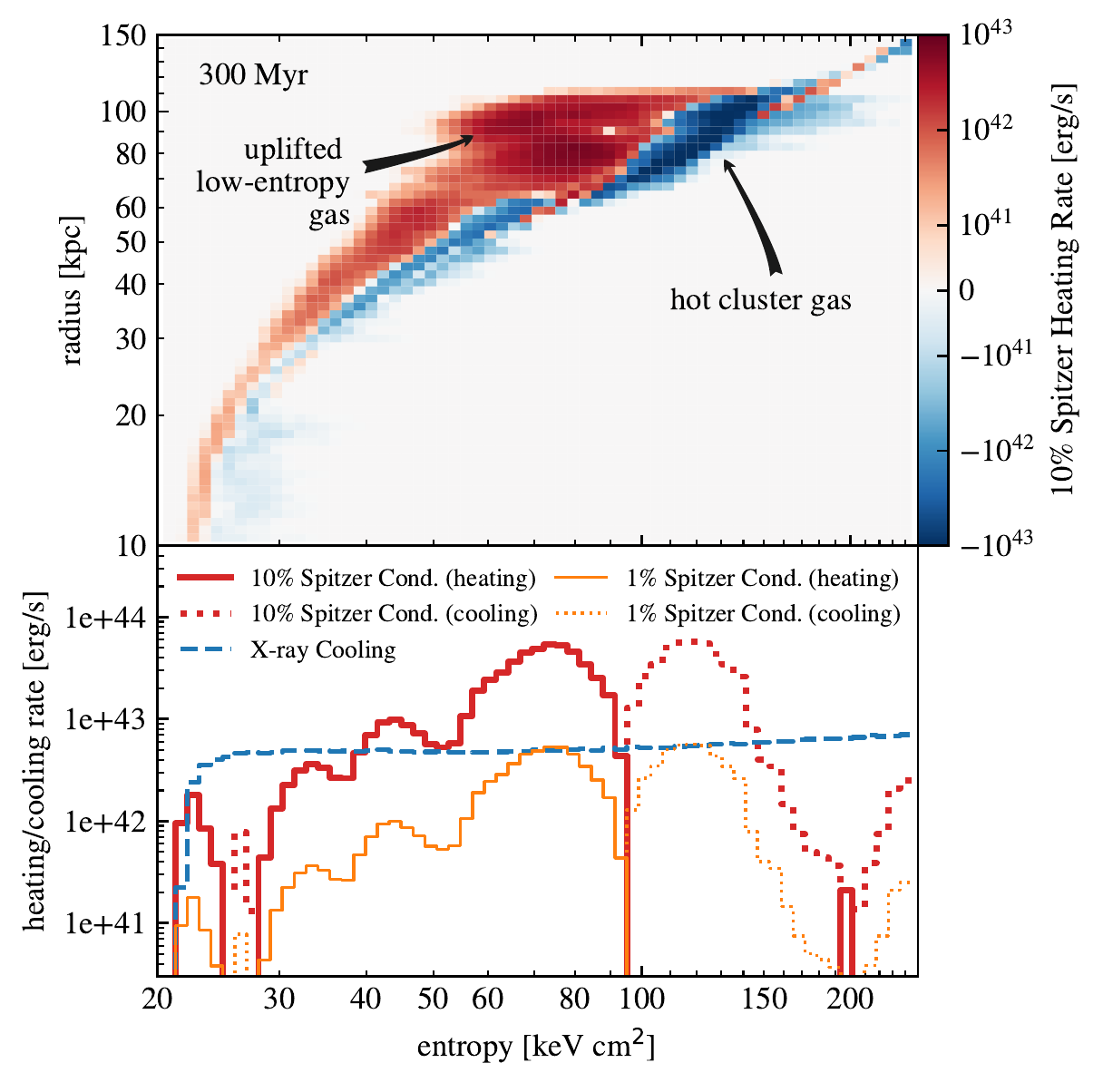}
	\caption{Same as Fig.~\ref{fig:heating_rate_100Myr} but at 300 Myr. The
		heating rates are large, especially for the entropy bins between 55 and
		95. These are the gas that were originally closer to the core and were
		brought to larger radii by the rising bubbles.}
	\label{fig:heating_rate_300Myr}
\end{center}
\end{figure}

The instantaneous global heating and cooling rate due to conduction is
calculated by taking the divergence of the heat flux. We then add the thermal
conduction heating/cooling rates in each radial and entropy bin and compare
them with the x-ray cooling rate in the marginal histogram at two different
times in Fig.~\ref{fig:heating_rate_100Myr} and
Fig.~\ref{fig:heating_rate_300Myr}. We show two plausible conduction rates at
10\% and 1\% of the Spitzer value. The heat flux in galaxy clusters is likely
far below the nominal Spitzer value, For example, a factor of 100 suppression
is consistent with recent results from studies of the plasma physics of
conduction
\citep{Roberg-Clark2018}.

One can see that at an earlier time in Fig.~\ref{fig:heating_rate_100Myr}, the
conductive heating rate at 10\% Spitzer value does not yet overcome the x-ray
cooling rate. At 300 Myr in Fig.~\ref{fig:heating_rate_300Myr}, the heating
rates can be close to the radiative cooling in some bins even at
$f_{\mathrm{Sp}}=0.01$. The heating and cooling rate as a function of radius
have much smaller values since most of the Spitzer heat flux happens between
the high-entropy and the low-entropy gas at the same radius as shown in the
upper panels in Fig.~\ref{fig:heating_rate_100Myr} and
Fig.~\ref{fig:heating_rate_300Myr}. Note that the heating and cooling rates in
these figures are cluster-average quantities. While x-ray cooling is a global
phenomenon, the conductive heating and cooling happen primarily around the
uplifted gas. Thus, the net heating rate locally at the surface of the
low-entropy gas is significant even at 1\% of the Spitzer rate.

\begin{figure}
\begin{center}
	\includegraphics[width=0.49\textwidth]{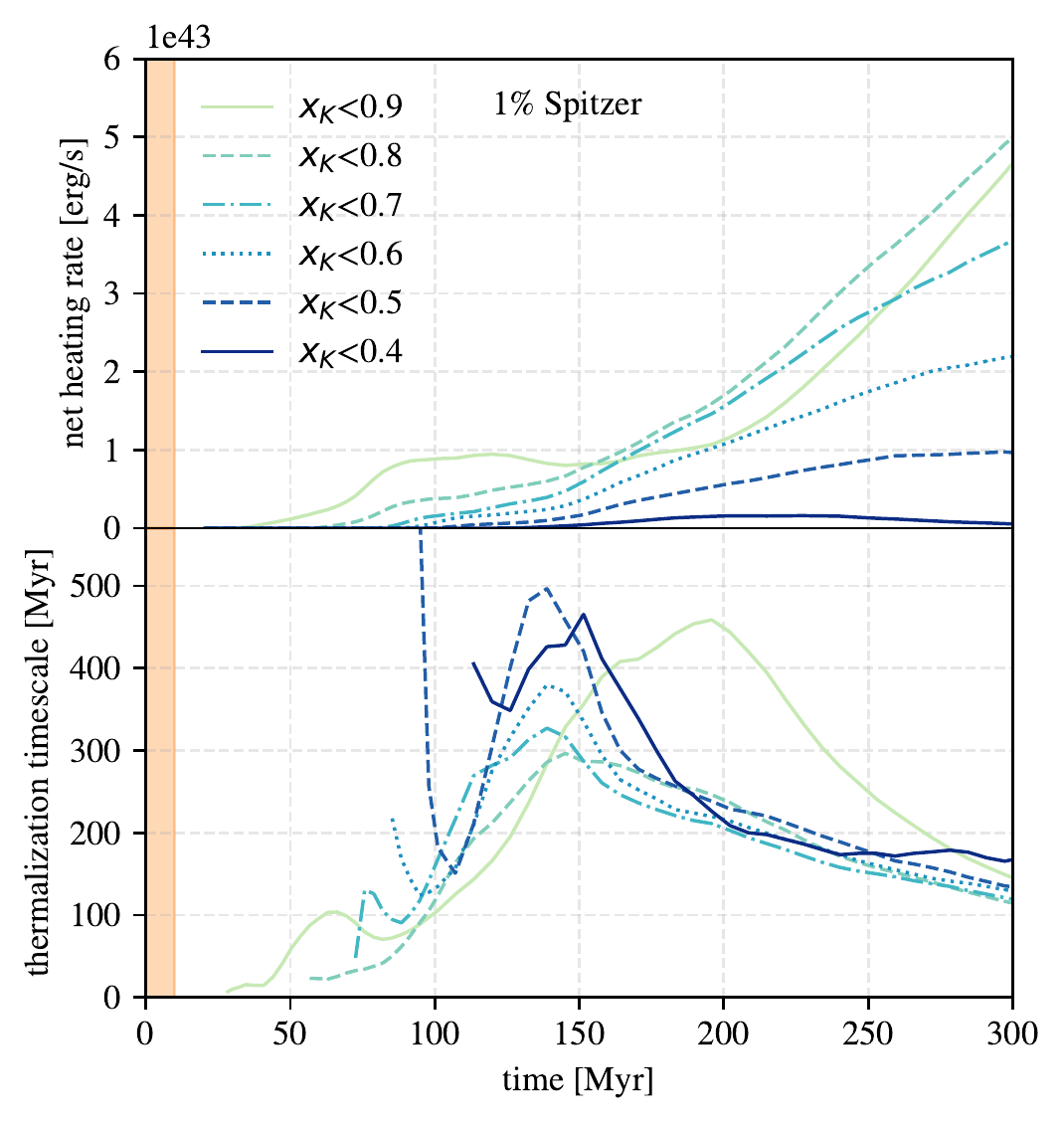}
	\caption{This figure shows the evolution of the aggregated net conductive
		heating rate at 1\% Spitzer value minus the x-ray cooling rate for the
		low-entropy gas with entropy ratio $x_K$ (\emph{upper panel}), and the
		average heating timescale $\tau_{\text{heat}}$ to thermalize the energy
		budget of the low-entropy gas (cf. Fig.~\ref{fig:mass_energy_budget})
		to the surrounding temperature (\emph{lower panel}).}
	\label{fig:conduction_rate_timescale}
\end{center}
\end{figure}

Although the microphysical details of the conduction deserve investigations in
more depth, we find that thermal conduction will be able to bring the
low-entropy gas to the temperature of the heat reservoir in a short amount of
time. The heating rate increases significantly at later times, which is shown
in Fig.~\ref{fig:conduction_rate_timescale}, where we aggregate the net heating
rate of the low-entropy gas for 1\% Spitzer value and subtract the x-ray
cooling rate. Most of the lower-entropy-ratio gas experiences net heating.
However, at later times, cooling exceeds heating in the gas with entropy ratio
between 0.8 and 0.9, which is indicated by the cumulative heating curve for
$x<0.9$ crossing below the $x<0.8$ curve after 150 Myr.

We can also estimate the heating timescale $\tau_{\text{heat}}$ from the energy
budget and the corresponding net heating rate (lower panel of
Fig.~\ref{fig:conduction_rate_timescale}). The free-fall time
$\tau_{\text{ff}}$ at 100 kpc, where most of the low-entropy gas is located at
the simulation time of 300 Myr, is about 150 Myr. At 1\% of the Spitzer rate,
the thermalization timescales are comparable or shorter than the free-fall
time. As long as the lifted gas is thermalized before it falls back, the heat
pump mechanism will be able to pull more energy than the AGN provided. In the
discussion section, we will discuss the conduction rates in more details.

We perform the same calculation on a similar simulation with different jet
magnetic field configuration and find the same conclusion. Locally, the
conduction rate is slightly different due to different bubble shapes, but
globally the energy budget remains much the same within 10 per cent.

\section{The Maximum Possible Efficiency of the Cluster Heat Pump}
\label{sec:efficiency}
In a traditional heat pump, energy is expended on performing mechanical work
(compression or decompression) before the heat exchange stage. In geothermal
heat pumps, energy is expended to move fluid against gravity. Our case is
similar to a geothermal system in that work is done against net gravity.

The low-entropy gas is lifted against the net gravitational force towards the
heat reservoir in order for thermal conduction to take place. Due to the
buoyancy, lifting the gas from its original location, where the density is the
same as the surroundings, requires almost no force. Once the gas is away from
its original location, the density drops adiabatically. As long as the density
profile of the cluster falls steeper than adiabatic, i.e. the ICM profile is
sub-adiabatic, which is always true for convectively stable cool-core clusters,
lifting the gas requires work. In this case, the temperature of the lifted gas
is lower than the surroundings and allows thermal energy to transfer to the gas
through conduction.

\begin{figure*}
\begin{minipage}{1\textwidth}
	\includegraphics[width=1.0\textwidth]{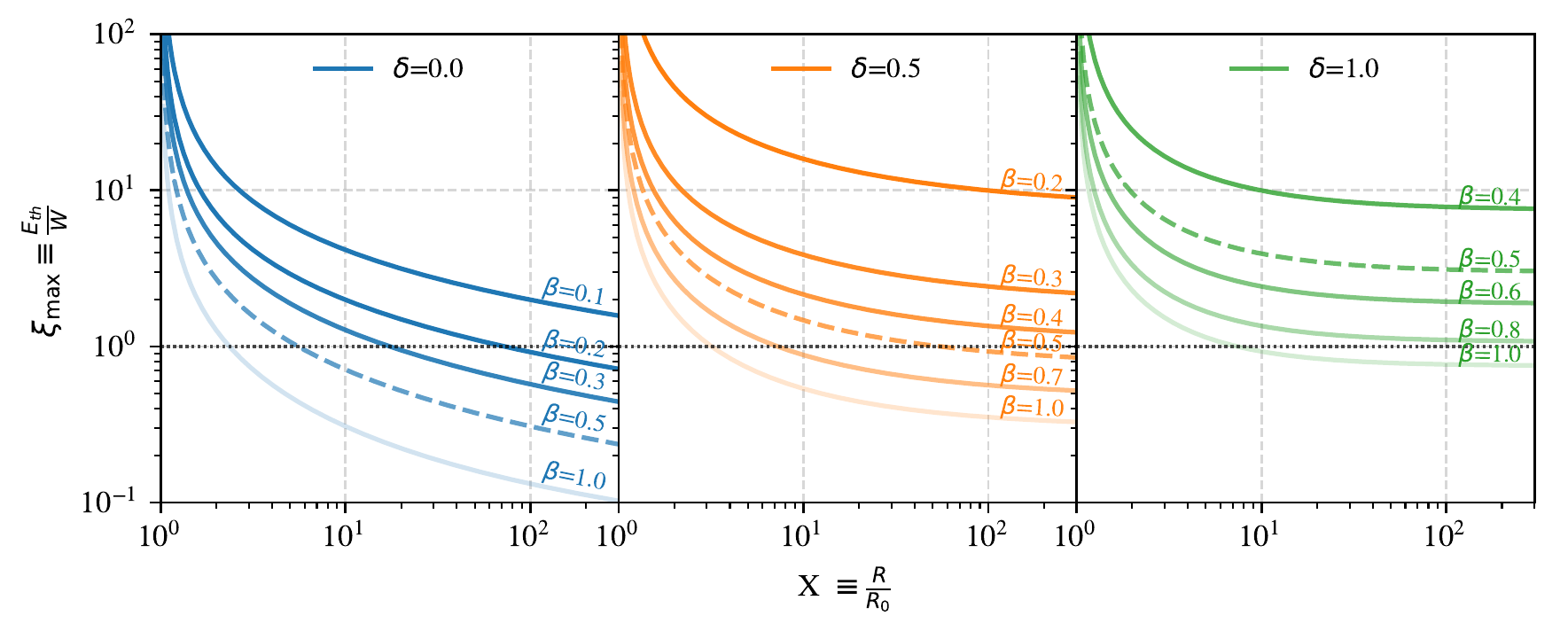}
	\caption{Efficiency for various $\beta$ and $\delta$ (as defined in
		Eq.~\ref{eq:power_laws}) for monatomic gas. $\delta$=0 corresponds to
		an isothermal cluster. Note that $\beta$ must be larger $\delta/3$,
		otherwise the pressure gradient becomes positive. Dashed lines
		correspond to $\beta=0.5$ are most similar to the density profile used
		in our simulation. The efficiency larger than unity (dotted line) means
		that more thermal energy is transferred than the work needed to lift
		the gas. Note that high efficiency does not necessarily mean that the
		gas can gain a huge amount of thermal energy. See
		Fig.~\ref{fig:energy_gain} for the energy gain.}
	\label{fig:energy_efficiency}
\end{minipage}
\end{figure*}

Here we consider the limiting efficiency of the heat pump, defined as the ratio
between the thermal energy through conduction and the work needed to move the
gas to a large radius, in a simplified case. We first consider the generic case
where the temperature and density profiles of the ICM follow simple power-laws
\begin{equation}
	T(r) = T_0 \left( \frac{r}{R_0} \right)^{\delta}.
	\quad
	\rho(r) = \rho_0 \left( \frac{r}{R_0} \right)^{-3\beta},
	\label{eq:power_laws}
\end{equation}
where the index of the density corresponds to $\beta$ in the $\beta$-model
outside of the core in Eq.~\ref{eq:beta_model}. When $\delta = 0$, this
represents an isothermal atmosphere; for cool-core clusters, $\delta > 0$. The
pressure then follows the ideal gas law
\begin{equation}
	P(r) = P_0 \left( \frac{r}{R_0} \right)^{-3\beta+\delta},
\end{equation}
We also assume the ICM is in hydrostatic equilibrium
\begin{equation}
	\frac{\dd P}{\dd r} = -g\cdot\rho,
\end{equation}
which gives us the condition that $-3\beta+\delta<0$ for negative pressure
gradient and the scaling of the gravitational acceleration
\begin{equation}
	g(r) = g_0 \left( \frac{r}{R_0} \right)^{\delta-1}.
\end{equation}

\begin{figure*}
\begin{minipage}{1\textwidth}
	\includegraphics[width=1.0\textwidth]{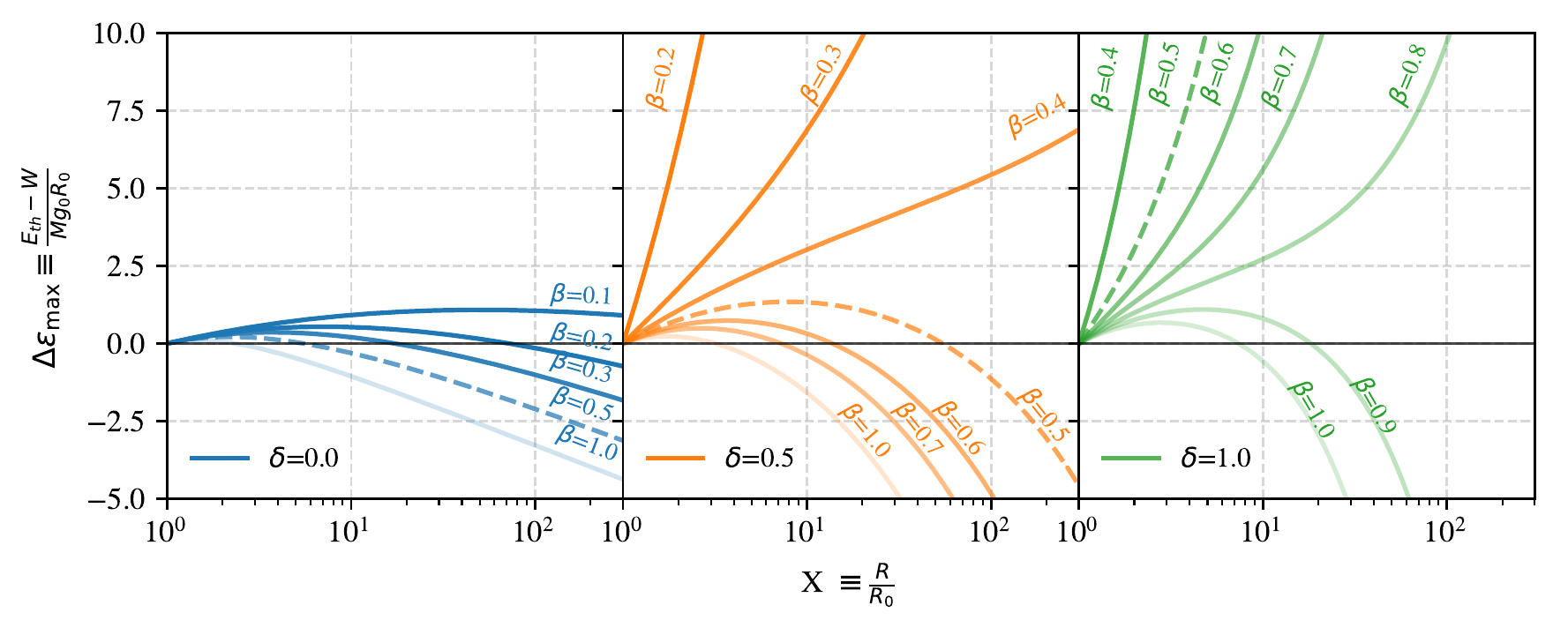}
	\caption{Energy gain for various $\beta$ and $\delta$ for monatomic gas.
		Dashed lines correspond to $\beta=0.5$ are most similar to the density
		profile used in our simulation. For $\beta < 5\delta/6 $ the curve
		diverges to infinity since lifting to larger radii will provide more
		energy; for $\beta > 5\delta/6 $, an optimal lift of $X$ can be found
		at the peaks of the curves that the heat pump acquires most thermal
		energy than the expenditure on work for lifting.}
	\label{fig:energy_gain}
\end{minipage}
\end{figure*}

In the limiting case of slow lift of a blob of gas, we can neglect the kinetic
energy of the blob. When the gas is lifted away from its original location, it
will expand to be in pressure balance with the surroundings. The net
gravitational force acting on the gas is
\begin{equation}
	F_g = -\Delta \rho ~V g = -(\rho_{\mathrm{blob}} - \rho(r)) V_{\mathrm{blob}}~g(r)
\end{equation}
Assuming the gas expands adiabatically, we have the scaling of the volume
\begin{equation}
	V_{\mathrm{ad}}(r) = V_0 \left( \frac{P(r)}{P_0} \right)^{-\frac{1}{\gamma}} = V_0 \left( \frac{r}{R_0} \right)^{\frac{3\beta-\delta}{\gamma}}
\end{equation}
and the density
\begin{equation}
	\rho_{\mathrm{ad}}(r) = \rho_0 \left( \frac{r}{R_0} \right)^{-\frac{3\beta-\delta}{\gamma}}.
\end{equation}
The net gravitational force can then be written as
\begin{equation}
\begin{aligned}
	F_g(r) &= -\left(\rho_{\mathrm{ad}}(r) - \rho(r)\right) V_{\mathrm{ad}}(r)~g(r)\\
	&= -M g_0 \left[ \left( \frac{r}{R_0}\right)^{\delta-1} - \left( \frac{r}{R_0}\right)^{-3\beta+\delta-1+\frac{3\beta-1}{\gamma}} \right] ,
\end{aligned}
\end{equation}
where $M$ is the mass of the lifted gas. The work needed to lift the gas from
$R_0$ to $R$ is the integral of the force
\begin{equation}
\label{eq:work}
\begin{aligned}
	W &= \int_{R_0}^{R} - F_g(r) \dd r\\
	&= M g_0 R_0\int_{1}^{X} x^{\delta-1}\left( 1 - x^{-3\beta+\frac{3\beta-\delta}{\gamma}} \right)\dd x,
\end{aligned}
\end{equation}
where we changed variables to $x \equiv r/R_0$ and call the lifting radius
ratio $X \equiv R/R_0$.

When the gas is lifted, its temperature is lower than the surroundings
\begin{equation}
	T_{\mathrm{ad}}(x) = T_0 \frac{ P(x) }{ \rho_{\mathrm{ad}}(x) }\frac{\rho_0}{P_0}
				  = T_0 ~x^{-3\beta+\delta+\frac{3\beta-\delta}{\gamma}}
\end{equation}
and the thermal energy necessary to themalize the gas when it is lifted to
$r=R$ is
\begin{equation}
\label{eq:thermal_energy}
\begin{aligned}
	E_{\mathrm{th}} &= \frac{1}{\gamma-1}\frac{M}{\mu m_p} k ~(T(X) - T_{\mathrm{ad}}(X))\\
	&= \frac{1}{\gamma-1}\frac{M}{\mu m_p} kT_0 \left( X^{\delta} - X^{-3\beta+\delta+\frac{3\beta-\delta}{\gamma}} \right)\\
	&= \frac{1}{(3\beta-\delta)(\gamma-1)}M g_0 R_0 X^{\delta} \left( 1 - X^{-3\beta+\frac{3\beta-\delta}{\gamma}} \right),
\end{aligned}
\end{equation}
where the hydrostatic equilibrium and the ideal gas law imply
\begin{equation}
	\frac{k T_0}{\mu m_p} = \frac{g_0 R_0}{3\beta-\delta}.
\end{equation}

The efficiency of the heat pump can then be expressed as
\begin{equation}
\begin{aligned}
	\xi_{\mathrm{max}} &\equiv \frac{E_{\mathrm{th}}}{W}\\
	&= \frac{1}{(3\beta-\delta)(\gamma-1)}\frac{X F(X)}{\int_{1}^{X} F(x) \dd x},
\end{aligned}
\end{equation}
where $F(x) \equiv x^{\delta-1}\left( 1 -
x^{-3\beta+\frac{3\beta-\delta}{\gamma}} \right)$. We plot the efficiency for
several cases of $\delta$ and $\beta$ in Fig.~\ref{fig:energy_efficiency}.
Note that the efficiency is largest at small $X$ because of the near-zero
work needed to lift the gas around its original location. The gravitational
energy is invested in lifting the gas upward and will not be recovered
if the gas is thermalized before it sinks back.
In the case of large lift, i.e. $X \gg 1$, the limiting efficiency becomes
\begin{equation}
	\lim_{X \gg 1} \xi_{\mathrm{max}} = \frac{\delta}{(3\beta-\delta)(\gamma-1)}.
\end{equation}
For monatomic gas, $\gamma = 5/3$, the limiting efficiency is larger than 100
per cent for large $X$ when
\begin{equation}
	\beta < \frac{5}{6}\delta.
\end{equation}
In Fig.~\ref{fig:energy_efficiency}, we can see that the heat pump mechanism
operates optimally when the temperature gradient is large (large $\delta$) or
when the density gradient is small (small $\beta$) such that lifting the gas
requires very little work. Although the efficiency is larger at small $X$,
the absolute thermal energy gain is small. We can further consider the
energy gain of the mechanism normalized by the initial gravitational energy
\begin{equation}
\begin{aligned}
	\Delta \epsilon_{\mathrm{max}} &\equiv \frac{E_{\mathrm{th}}-W}{M g_0 R_0}\\
	&= \frac{1}{(3\beta-\delta)(\gamma-1)} X F(X)-\int_{1}^{X} F(x) \dd x.
\end{aligned}
\end{equation}
We show the energy gain for various $\delta$ and $\beta$ in
Fig.~\ref{fig:energy_gain}. For $\beta < 5\delta/6$, the curves diverge to
infinity, i.e. more thermal energy is available for more lift. For $\beta >
5\delta/6 $, there is an optimal lift $X$ that maximize the energy gain.
Considering the temperature gradient of the cluster is only positive from the
cool core to the hot atmosphere, it is unlikely that this mechanism will get
infinite thermal energy from large radial change. Instead, there will be an
optimal lift that the energy gain is maximized depending on the profile
of the cluster.

Since the AGN expends energy on other aspects not included here (namely, the
kinetic and internal energy of the jets/lobes, and kinetic energy of the lifted
gas), the true efficiency is $\xi < \xi_{\mathrm{max}}$ and the energy gain is
$\Delta \epsilon < \Delta \epsilon_{\mathrm{max}}$. This method works as long
as positive heat exchange is possible, i.e. the temperature gradient being
shallower than the adiabatic gradient, $\delta > 3\beta(1-\gamma)$, which is
the criterion for convective stability.

\begin{figure}
\begin{center}
	\includegraphics[width=0.49\textwidth]{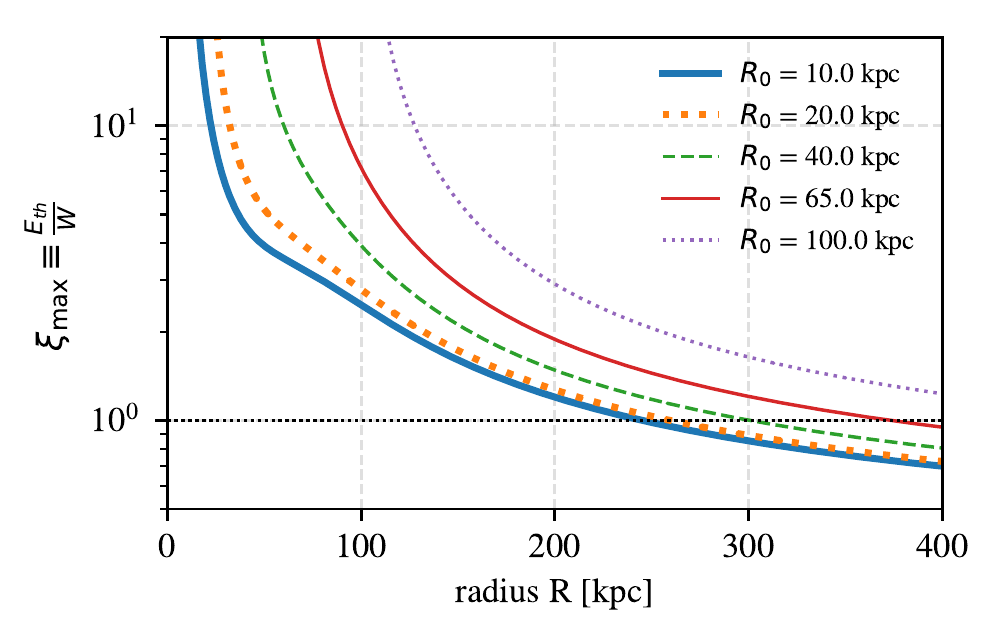}
	\caption{Efficiency for various initial radii $R_0$ for the ICM profiles
		used in our simulation that represent the Perseus cluster. Since the
		temperature tends to isothermal at large radii, the efficiency is not
		larger than 100 per cent asymptotically. Note that the AGN bubbles rise
		only up to $\sim$ 100 kpc in our simulation, which falls in the very
		high efficiency range.}
	\label{fig:energy_efficiency_perseus}
\end{center}
\end{figure}

\begin{figure}
\begin{center}
	\includegraphics[width=0.49\textwidth]{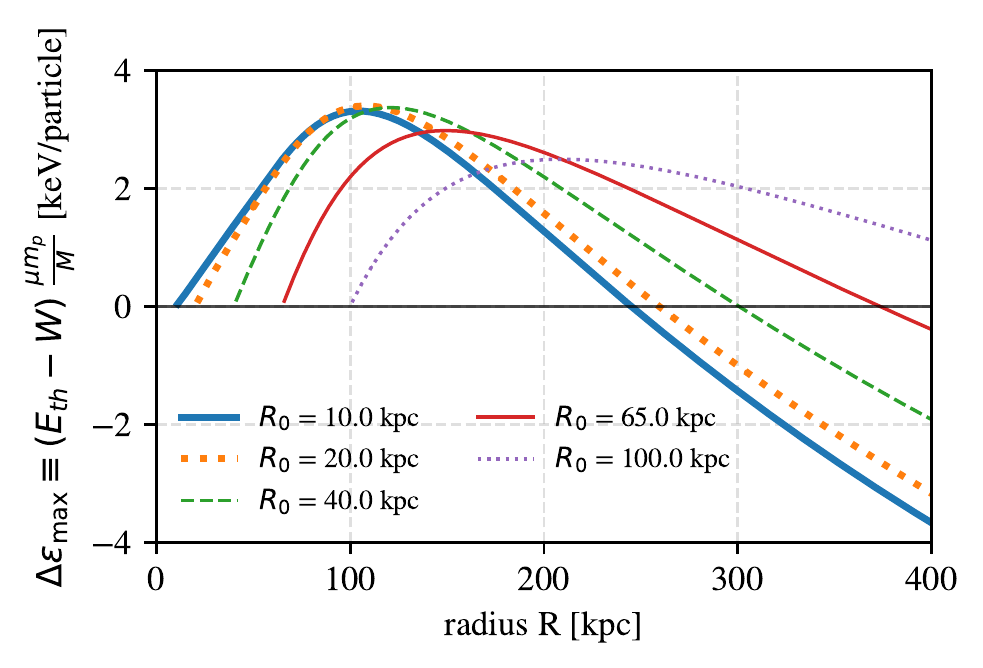}
	\caption{Average energy gain per particle for various initial radii $R_0$
		for the ICM profiles used in our simulation that represent the Perseus
		cluster. This is the extra thermal energy a particle can get when
		it is lifted.}
	\label{fig:energy_gain_perseus}
\end{center}
\end{figure}

For a more realistic background profile like the one used in our simulation of
the Perseus cluster (sec Section \ref{sec:numerical_techniques}), we can
integrate the work and calculate the conductive thermal energy numerically,
which is shown in Fig.~\ref{fig:energy_efficiency_perseus} for efficiency and
Fig.~\ref{fig:energy_gain_perseus} for energy gain. Various initial
radii are included in the calculation. We can see that a lift of a few tens of
kpc, as observed with the simulations, can provide for a high limiting
efficiency and a large energy gain. This radius corresponds to the final
distance of the bubbles from the cluster center in our simulation, as well as
the locations of x-ray bubbles in many observations \citep{Birzan2004,
Shin2016}.

In Fig.~\ref{fig:energy_gain_perseus}, we can see lifting the low-entropy gas
can provide up to an additional $\sim$ 3 keV of energy per particle. Thus,
moving $10^{10} M_{\odot}$ can generate additional $10^{59}$ erg of thermal
energy from the heat pump mechanism as we estimate from the simulation in
Section \ref{subsec:mass_energy_budget}.

\begin{figure}
\begin{center}
	\includegraphics[width=0.49\textwidth]{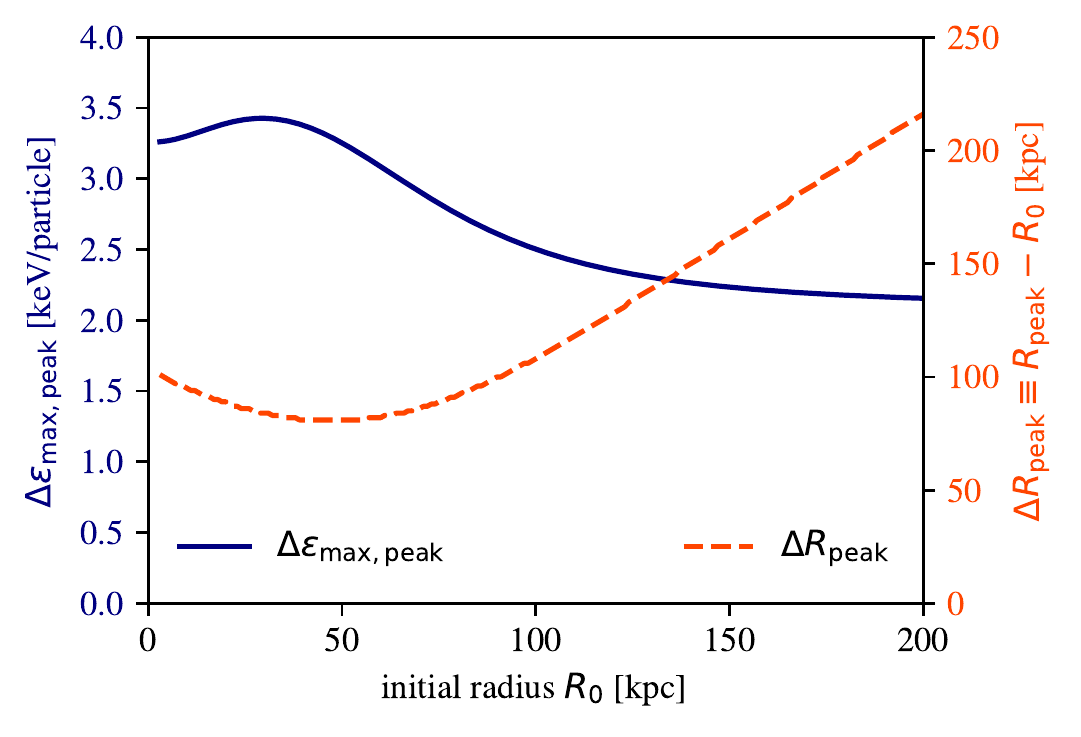}
	\caption{Peak energy gain per particle and the lifting distance needed.
		This figure shows the peak of the possible energy in
		Fig.~\ref{fig:energy_gain_perseus} and the corresponding change in
		radius.}
	\label{fig:energy_gain_peak_perseus}
\end{center}
\end{figure}

Now we can find the peak energy gain for different initial radii in
Fig.~\ref{fig:energy_gain_perseus} and the corresponding change in radius,
which is shown in Fig.~\ref{fig:energy_gain_peak_perseus}. We see that for
small initial radius, the peak energy gain is larger and also requires less
lifting. For large initial radius, where the temperature profile tends to be
isothermal, the $\Delta R_{\mathrm{peak}}$ increases linearly, i.e. constant
$X_{\mathrm{peak}}$, and the energy gain becomes constant, i.e. constant $g_0
R_0$. Again, we see that the optimal operating range of the heat pump mechanism
is to lift the gas from the cool core to the hot atmosphere of the cluster.

For other observed groups and galaxies, the best fit $\beta$-model usually
gives $\beta$ in the range between 0.3 and 0.7 \citep{Mulchaey2003,
McCarthy2004, Dong2010, Osullivan2017}, including the Perseus Cluster modeled
in our simulation. Most of their temperature profiles do not follow a simple
power law, but increase from the cool core to the outskirt and becomes near
isothermal at large radii. Thus, the heat pump operates best across the
temperature gradient around the cluster core where the heating is essential for
the cluster.

\section{Discussion}
\label{sec:discussion}

\subsection{The conduction rate and energy budget}

In the literature, adiabatic uplift is usually deemed ineffective to offset the
cooling because the lifting of gas only decreases the cooling time by a factor
of few \citep[see][for review]{McNamara2012}. However, in our scenario, even
though the uplift process is adiabatic, the gas will be heated by the hot
atmosphere of the cluster due to the increased conduction coefficient and
temperature gradient. The thermalization timescale based on 1\% Spitzer value
in Fig.~\ref{fig:conduction_rate_timescale} suggests that the low-entropy gas
might not have enough time to be heated during the uplift, but will eventually
be heated before sinking. The energy budget estimated from the simulation in
Fig.~\ref{fig:mass_energy_budget} provides a rough estimation of the available
energy through this mechanism. The actual energy budget depends on various
factors, including the conduction rate, the efficiency of lifting, and the
cluster profile. We show in this work that this effect can be significant and
should not be ignored when analyzing the thermal balance of clusters.

Anisotropic thermal conduction has been widely discussed in the context of
galaxy clusters. \cite{Kannan2017} pointed out that anisotropic thermal
conduction could enhance the energy coupling between the AGN and the ICM
through more effective mixing. In this work, we focus on the additional
heating channel between high- and low-entropy cluster gas facilitated by the
long-lived rising bubbles, rather than the direct heating from the bubbles
themselves.

We caution that the calculation in Section \ref{subsec:spitzer_heating_rates}
is likely an overestimate due to the following reasons: (a) Our simulation does
not have explicit thermal conduction, and the estimated rate is based on the
distribution of the gas without prior heat exchange (except for numerical
mixing). Once thermal conduction takes place, the sharp temperature gradient
will quickly smooth out and the rate will drop roughly as $t^{1/2}$. (b) Heat
flux might be significantly suppressed by magnetic field geometry and
microscopic plasma instabilities in the cluster gas \citep{Roberg-Clark2016,
Roberg-Clark2018a}. Although we include magnetic fields in the injection of
the jets, the ICM is not magnetized in our simulations. Thus, it is impossible
to estimate the effect of anisotropic conduction between the low-entropy gas
and the hot atmosphere.

However, we argue that these uncertainties do not change our major finding that
the heat pump mechanism is able to provide additional energy to the AGN
feedback loop. The total energy budget estimated in Section
\ref{subsec:mass_energy_budget} and the efficiency in Section
\ref{sec:efficiency} should not change significantly due to the uncertainties
in the conduction rate provided that the thermalization can happen before the
sinking of the low-entropy gas. As long as the heating timescale is shorter
than the free-fall time ($\tau_{\text{heat}} < \tau_{\text{ff}}$), our
mechanism works, which leaves a lot of room even if the conduction coefficient
is much smaller than Spitzer. We will leave the investigation of the
anisotropic conduction effect and the interplay between the conduction rate and
the total energy budget to future work.

Even if the conduction rate at an earlier time is substantial and the gas is
thermalized before it reaches the largest radius, the upward motion triggered
by the rising bubble can still bring the already thermalized gas to larger
radii. Indeed, we show in Section \ref{sec:efficiency} that the heat pump
operates at very high efficiency around the initial radius (X $\sim$ 1). This
is due to the minimal work needed to lift the gas with the same density as the
surroundings. Thus, the bubble might be able to bring more gas to large radii
if the gas was thermalized before reaching the final location.

The amount of the uplifted low-entropy gas, nevertheless, might change
depending on the integrity of the bubble while it rises. Pure hydrodynamic
bubbles have been suggested to be quickly disrupted by instabilities
\citep{Diehl2008a}. Thus, most of the bubbles associated with x-ray cavities
may be supported by magnetic fields or require other mechanisms to suppress
the instabilities. Some research suggests that the stability of the bubbles
might be affected more by the surrounding magnetic field than by the field
inside the bubbles \citep{Ruszkowski2007, Dursi2008}. How the amount of lifted
low-entropy gas depends on the dynamics of the bubble and the internal or
external field, will require further research.

\subsection{Effect of mixing}

Some recent studies identify mixing (between the bubble plasma and the cluster
gas) as the primary heating mechanism \citep{ Yang2016, Li2017}. Whether mixing
is a major mechanism requires further studies \citep[see][]{Hillel2016,
Hillel2018}. It is important to note, however, that the effect of the thermal
conduction is only evident in simulations with high spatial resolution. In
resolution-limited simulations, both the hot gas from the bubbles and the low
entropy gas lifted in their wake is quickly mixed with the ICM and unrealistic
numerical mixing leads to quick dissipation of the bubble and increased heating
of low-entropy uplifted gas. The amount of lifted low-entropy gas is likely
reduced due to the bubbles being prematurely mixed and dispersed numerically.
High resolution is important also because conduction is strongly increased by
the corrugated surface of the uplifted gas and turbulence keeps re-arranging
the interface between low- and high- temperature gas, regenerating the hot-cold
interface and keeping gradients large for large heating rate (see
Fig.\ref{fig:entropy_ratio}).

On the other hand, our simulation probably still underestimates the level of
turbulent mixing due to numerical viscosity. Such mixing between low-entropy
and hot cluster gas, however, would help to bring cold and hot gas into contact
and therefore increase the heat transfer rate. This strengthens our argument
that the low-entropy gas can be heated before falling back to the core.

\subsection{Multiphase filaments}

Multiphase filaments in H$\alpha$, x-ray, and molecular lines, are often
observed in galaxy clusters \citep[see e.g.][]{Fabian2003a, Lim2012,
McNamara2014}. Some observations \citep{Vantyghem2016, Russell2016,
Vantyghem2018} and simulations \citep{Revaz2008, Li2014} suggest
that the filaments are formed \emph{in-situ} from the entrained cold gas lifted
by the bubbles. Our simulations do not include the formation of the filaments
as the entropy reservoir of the uplifted gas in our simulations corresponds to
a different ICM phase. The magnetic field in the wake might indeed prevent
evaporation by thermal conduction of the H$\alpha$ filaments
\citep{Ruszkowski2007}. However, the existence of the filaments may be
suggestive of efficient up-lift. Regardless, there is no way to generally avoid
uplift of cold gas by jet-inflated bubbles.

While our results suggest that the conduction rate between the hot atmosphere
and the lifted low-entropy gas could be substantial, our simulations do not
exclude the formation of the filaments since they involve more physics that
were not modeled in our work. The H$\alpha$ filaments could be a much smaller
volume where thermal instability takes place and thus consistent with the
stimulated condensation scenario \citep{McNamara2016}. Locally, cooling can
become catastrophic in this scenario, as a detailed local balance between
heating and cooling is not provided by the heat pump mechanism, which regulates
the cluster on a global scale. Rapidly cooling gas clumps will sink to the
gravitational center of the cluster, where they are most easily captured by
the drag of the rising bubbles.

\subsection{Implication of the heat pump mechanism}

In our simplified simulation, we study the long-term effect of a single AGN
outburst. The proper balance between heating and cooling is not the main goal
of this work since it is impossible to counteract the cooling of the cluster
for 300 Myr with just one episode of AGN activity. It has been a question how
the ``bursty'' AGN events can regulate the constant radiative cooling
\citep{Best2007}. The process discussed in this work suggests a solution to
this problem by transforming the bursty nature of AGN into a much more gentle
heating process. Not only can the AGN heat the ICM during the active phase, but
the rising bubbles enable conduction between the hot atmosphere and the
uplifted cool gas. It can also help to explain the scatter of the jet power
seen in the observations relative to the cluster cooling power since this
mechanism does not require a close instantaneous balance between jet power and
the cooling rate.

By bringing gas of different entropy into close thermal contact, the AGN
essentially acts as a heat pump. The energy used to heat the low-entropy gas is
drawn from the thermal base of the outer cluster, not the jet power. In this
way, the heating efficiency of AGN can exceed 100 per cent, as slow and
subsonic exchange of gas can be done adiabatically. We show that the thermal
energy in many cases exceeds the work needed to bring the low-entropy gas to
large radii in Section \ref{sec:efficiency}.

The feedback loop of the heat pump is, like the traditional AGN feedback,
governed by the cooling within the cluster core that stimulates the AGN
activities. In the short term, other mechanisms may still be required to help
counteract the rapid cooling in the core if there were no previous episodes of
AGN bubbles. Once the rising bubbles are established, they will be able to
remove part of the cooled gas from the core and regulate the cooling in a long
timescale and with more energy than injected by the jets itself. This will
likely change the necessary duty cycle to balance the radiative cooling in AGN
feedback.

\section{Conclusions}
\label{sec:conclusions}

We consider the idea that AGN jets could act as a ``\emph{heat pump}'' by
inflating buoyant bubbles that lift low-entropy gas from the core into thermal
contact with the surrounding hot gas. The increased temperature gradients and
conduction coefficient accelerate the thermal exchange between the hot
atmosphere and the lifted gas. The AGN in this scenario does not heat the
cluster core directly; rather it creates a pipeline which allows the energy
exchange between the heat reservoir and the cool core. In this mechanism, the
total available energy gained by the cluster core to offset the radiative
cooling is not simply limited by the total energy output of the AGN, but also
the heat transferred from the hot gas at large radii of the cluster to the
uplifted low-entropy gas. Our analysis implies that a 10-Myr active jet could
still affect the thermal state of the cluster after 300 Myr. This mechanism has
the advantage that bursty AGN activity is transformed into a more gentle and
longer-lasting heating process. In this work, we demonstrate that
\begin{enumerate}
\item Jet-inflated bubbles can bring a significant amount of low-entropy gas to
	large radii. The removal of the cool gas from the core is seen mostly in
	the wake of the rising bubbles. Higher-entropy gas replaces the lifted
	low-entropy gas. Repeated AGN cycles can thus likely induce sufficient
	circulation to affect gas not aligned with the jet axis.

\item The energy budget that can be drawn from the heat reservoir to the
	low-entropy gas is comparable to the energy input from the jet. This energy
	budget reaches a maximum when the bubbles are already mostly disrupted.

\item The large thermal conduction rate at later times suggests low-entropy gas
	can be thermalized before sinking back even at 1\% of Spitzer conduction
	rate. Uncertainties and open questions regarding the nature of thermal
	transport in the intracluster plasma motivate further work to explore this
	proposed mechanism.

\item The efficiency of the heat pump, defined as the ratio between the thermal
	energy transferred and the work needed to lift the low-entropy gas, could
	be greater than 100 per cent in many cases. Even if the efficiency is not
	larger than 100 per cent asymptotically, the energy gain will still be
	positive for moderate lift and will work in concert with any other forms of
	dissipation of jet energy.

\end{enumerate}

We want to bring attention to the heat exchange between the uplifted gas and
the hot atmosphere through this exploratory work inspired by the long-term
evolution of our simplified numerical experiment. In addition to heating the
core itself, the AGN could act as a facilitator in cool-core systems and
exchange energy from the hot atmosphere. To investigate this heat-pump
mechanism rigorously, we will need the simulations to include effects like
radiative cooling and anisotropic thermal conduction and also consider the
integrity of the bubbles. Our preliminary analysis nonetheless indicates that
this process can provide another avenue in the already hotly debated cooling
flow problem.

\label{lastpage}
\section*{Acknowledgement}
We thank the anonymous referee's insightful and detailed feedback that greatly
improves this work. YC and SH would like to acknowledge support from NASA
through the Astrophysics Theory Program grant NNX17AJ98G. This work used the
Extreme Science and Engineering Discovery Environment (XSEDE) Stampede at the
Texas Advanced Computing Center at The University of Texas at Austin and the
HPC Cluster at the Center for High Throughput Computing at the University of
Wisconsin-Madison. Support for this research was provided by the Office of the
Vice Chancellor for Research and Graduate Education at the University of
Wisconsin-Madison with funding from the Wisconsin Alumni Research Foundation. A
part of this work was presented in IAU Symposium 342: \textit{Perseus in
Sicily: from black hole to cluster outskirts} and was included in the
Proceedings.

\addcontentsline{toc}{section}{Bibliography}
\bibliographystyle{mnras}
\bibliography{./bibliography}

\end{document}